\documentclass[11pt]{article}

\usepackage[final]{acl}

\usepackage{times}
\usepackage{latexsym}

\usepackage[T1]{fontenc}

\usepackage[utf8]{inputenc}

\usepackage{microtype}

\usepackage{inconsolata}

\usepackage{graphicx}

\usepackage{amsmath}
\usepackage{amsfonts}
\usepackage[most]{tcolorbox}
\usepackage[table]{xcolor}
\usepackage{enumitem}
\usepackage{booktabs}
\usepackage{subcaption}

\title{Rethinking LLM Watermark Detection in Black-Box Settings: A Non-Intrusive Third-Party Framework}

\author{
 \textbf{Zhuoshang Wang\textsuperscript{1,2}},
 \textbf{Yubing Ren\textsuperscript{1,2}\thanks{Corresponding author.}},
 \textbf{Yanan Cao\textsuperscript{1,2}},
\\
 \textbf{Fang Fang\textsuperscript{1,2}},
 \textbf{Xiaoxue Li\textsuperscript{3}},
 \textbf{Li Guo\textsuperscript{1,2}}
\\
 \textsuperscript{1}Institute of Information Engineering, Chinese Academy of Sciences, Beijing, China \\
 \textsuperscript{2}School of Cyber Security, University of Chinese Academy of Sciences, Beijing, China \\
 \textsuperscript{3}National Computer Network Emergency Response Technical Team,\\Coordination Center of China (CNCERT/CC), Beijing, China
\\
\texttt{\{wangzhuoshang, renyubing\}@iie.ac.cn}
}

\begin{document}
\maketitle
\begin{abstract}
While watermarking serves as a critical mechanism for LLM provenance, existing secret-key schemes tightly couple detection with injection, requiring access to keys or provider-side scheme-specific detectors for verification. This dependency creates a fundamental barrier for real-world governance, as independent auditing becomes impossible without compromising model security or relying on the opaque claims of service providers. To resolve this dilemma, we introduce TTP-Detect, a pioneering black-box framework designed for non-intrusive, third-party watermark verification. By decoupling detection from injection, TTP-Detect reframes verification as a relative hypothesis testing problem. It employs a proxy model to amplify watermark-relevant signals and a suite of complementary relative measurements to assess the alignment of the query text with watermarked distributions. Extensive experiments across representative watermarking schemes, datasets and models demonstrate that TTP-Detect achieves superior detection performance and robustness against diverse attacks.
\end{abstract}

\section{Introduction}
Large language models (LLMs) have rapidly shifted from research prototypes to widely deployed generative systems, producing fluent text at negligible marginal cost. This unprecedented scalability has amplified long-standing concerns around synthetic content, as LLMs can now be used to automate misinformation campaigns, fabricate credible documents, and violate intellectual property rights. In parallel with these risks, LLM watermarking has emerged as a practical mechanism for provenance: by embedding subtle statistical signals during generation, watermarks enable LLM service providers to verify whether a suspicious text is watermarked, offering a promising solution for the proactive forensics of AI-generated content.

Most LLM watermarking techniques, regardless of whether they modify token probabilities \cite{pmlr-v202-kirchenbauer23a} or influence the sampling procedure \cite{dathathri2024scalable}, ultimately function by introducing a key-dependent bias toward certain tokens during generation. Then the detector must reconstruct this bias to verify the watermark; consequently, watermark injection and detection necessarily share the same secret key. This coupling poses a fundamental obstacle in real-world attribution. A watermark cannot be independently verified, since courts or platform moderators lack access to the key and must accept the service provider’s claim about whether a watermark is present. The process becomes opaque and trust-dependent. Allowing third parties to perform detection would require disclosing the key, which would compromise security by enabling adversaries to imitate or remove the watermark. Thus, current private-key schemes cannot simultaneously support independent verification and preserve the confidentiality on which their security depends.

This limitation makes a \textbf{Trusted Third Party (TTP)} essential for watermark governance: an entity that can verify the watermark without access to the secret key, preventing LLMs providers from misreporting results, blocking adversarial forgery, and enabling evidence suitable for regulatory or judicial use. Recent efforts toward third-party or publicly verifiable watermarking \cite{liu2024an,fairoze2025publiclydetectablewatermarkinglanguagemodels,duan2025pvmarkenablingpublicverifiability} take steps in this direction, but they retain a critical structural assumption: watermark injection and detection remain tightly paired, with verification logic tailored to specific injection mechanisms. This one-to-one coupling conflicts with real-world governance, where watermark injection is the responsibility of model providers, while verification and oversight belong to independent regulators. As long as detection is bound to the injection design, it cannot function as a neutral, reusable auditing layer across heterogeneous models and watermarking schemes. To resolve this mismatch, we introduce \textbf{TTP-Detect}, a key-agnostic black-box watermark detection framework that explicitly decouples detection from injection, allowing a TTP to assess watermark presence directly from output text alone.

TTP-Detect reframes absolute-threshold detection as a \textit{relative hypothesis testing} problem: given a query text, the goal is to determine whether it aligns more closely with the statistical behavior of watermarked outputs than with ordinary text. Our key observation is that no single statistic can reliably characterize watermark presence across algorithms, since different watermarking schemes concentrate their effects on different aspects of the generation process. TTP-Detect operationalizes this insight by first mapping texts into a representation space that amplifies \textit{watermark-relevant discrepancies}, and then evaluating the query through \textit{complementary relative measurements} that capture local consistency, global distributional shifts, and generation-time likelihood patterns (captured by post-hoc re-scoring). These signals provide partially independent evidence and are robustly calibrated into a \textit{unified decision score}. This design yields a detection framework that is robust across diverse watermarking schemes evaluated in our study and readily extensible via additional relative measurement modules.

Our main contributions are as follows:
\textbf{(1)} We introduce a pioneering formulation for third-party, key-agnostic watermark verification in black-box settings. By decoupling verification from injection, we establish a key-agnostic paradigm that enables third-party auditing without compromising security. \textbf{(2)} We propose TTP-Detect, a unified framework that reframes verification as a relative hypothesis test. By leveraging proxy representations and complementary relative measurements, our approach captures subtle watermark traces without access to the service provider's model states or specific algorithms. \textbf{(3)} We demonstrate superior universality and robustness through extensive experiments. TTP-Detect reliably identifies a wide range of representative watermarking schemes and maintains robustness under attacks.


\section{Related Work}
\paragraph{Private-Key Based LLM Watermarking} Existing LLM watermarking schemes generally fall into two primary categories: logits-based and sampling-based. The pioneering logits-based method, KGW~\cite{pmlr-v202-kirchenbauer23a}, partitions the vocabulary using a hash of preceding tokens seeded with a secret key, biasing generation toward green list tokens. To improve robustness against text editing, subsequent works propose global vocabulary partitioning~\cite{zhao2024provable} or semantic-aware partitioning~\cite{liu2024a,ren-etal-2024-robust,he-etal-2024-watermarks}. To better preserve text quality, \citet{hu2024unbiased,chen-etal-2025-improved} introduce unbiased reweighting, while \citet{ren-etal-2024-subtle,10.5555/3692070.3692903,10.5555/3692070.3693308,lee-etal-2024-wrote,lu-etal-2024-entropy,wang-etal-2025-morphmark} dynamically adjust the vocabulary partition or watermark strength. Sampling-based methods, in contrast, influence token selection without altering logits: \citet{aaronson2023watermarking,pmlr-v247-christ24a,fu-etal-2024-gumbelsoft,kuditipudi2024robust,dathathri2024scalable,mao-etal-2025-watermarking} explore token-level sampling strategies, while \citet{hou-etal-2024-semstamp,hou-etal-2024-k,dabiriaghdam-wang-2025-simmark} focus on sentence-level schemes. Recent advances further optimize the robustness--quality trade-off by synergistically combining logits- and sampling-based modifications~\cite{wang-etal-2025-trade}.

\paragraph{Publicly Detectable Watermarking} To mitigate the dependence on private keys during detection, some works explore publicly detectable or publicly verifiable watermarking schemes. UPV~\cite{liu2024an} implements generation and detection using two neural networks that share common token embeddings. \citet{fairoze2025publiclydetectablewatermarkinglanguagemodels} design a publicly detectable scheme that combines digital signatures, error-correcting codes and rejection sampling to embed a cryptographically verifiable signature into generated text. PVMark~\cite{duan2025pvmarkenablingpublicverifiability} further makes existing secret-key schemes publicly verifiable by wrapping the detectors with zero-knowledge proofs and re-engineering representative schemes into ZK-friendly circuits. However, their verification logic remains tightly coupled to the injection mechanism and typically depends on provider-controlled training or shared parameters, limiting portability across watermarking schemes and undermining detector neutrality.

\section{Preliminary}
\label{sec:preliminary}

\subsection{LLM Generation}
Consider an autoregressive LLM $\mathcal{M}$ with vocabulary $\mathcal{V}$. Given a prompt $x$ and a sequence of preceding tokens $y_{<i}$, the model predicts the next token $y_i$ iteratively. At each step $i$, $\mathcal{M}$ maps the context to a logits vector $\boldsymbol{l}_i = \mathcal{M}(x, y_{<i}) \in \mathbb{R}^{|\mathcal{V}|}$. This vector is then normalized via softmax to yield a probability distribution $\boldsymbol{p}_i$. Finally, the token $y_i$ is selected from $\boldsymbol{p}_i$ using a sampling strategy, denoted as $y_i \sim \mathrm{Sample}(\boldsymbol{p}_i)$.

\subsection{LLM Watermarking}
A watermarking algorithm consists of a generator and a detector. The generator modifies the standard generation process to produce watermarked text using a private key $\xi$. 

\paragraph{Logits-based Watermarking.}
These methods inject the watermark signal by modifying the logits $\boldsymbol{l}_i$ before the softmax layer. A representative method is the KGW~\cite{pmlr-v202-kirchenbauer23a} scheme. At each step $i$, the vocabulary $\mathcal{V}$ is partitioned into a green list $\mathcal{G}_i$ and a red list using a hash function seeded with the previous context and the private key. A bias $\delta$ is added to the logits of tokens in $\mathcal{G}_i$:
\begin{equation}
    \tilde{\boldsymbol{l}}_i[v] = \begin{cases} 
    \boldsymbol{l}_i[v] + \delta, & \text{if } v \in \mathcal{G}_i \\
    \boldsymbol{l}_i[v], & \text{otherwise}
    \end{cases}.
\end{equation}
This bias encourages the model to select tokens from $\mathcal{G}_i$. During detection, the detector calculates a z-score based on the observed proportion of $\mathcal{G}_i$ tokens in the text. If the z-score exceeds a threshold, the text is deemed watermarked.

\paragraph{Sampling-based Watermarking.}
These methods embed the watermark during the sampling stage. For example, AAR~\cite{aaronson2023watermarking} employs a pseudo-random sequence generated by the key $\xi$. The sampling process is modified to prioritize tokens that align with the sequence. The detection process verifies whether the generated tokens correlate significantly with the pseudo-random sequence derived from the key.

\section{TTP-Detect: A Black-Box Third-Party Detection Framework}

\begin{figure*}[t]
  \includegraphics[trim={0cm 6.6cm 0cm 4.1cm}, clip, width=\linewidth]{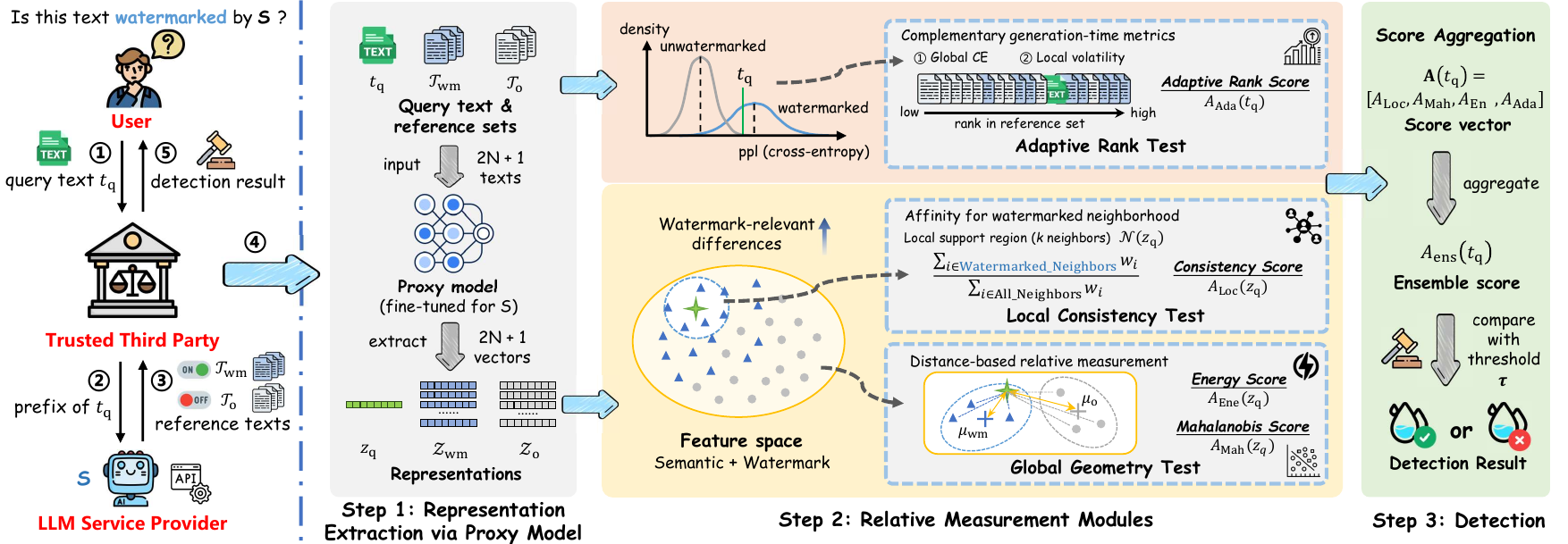}
  \caption {TTP-Detect framework overview.}
\end{figure*}

We first outline the TTP-Detect setting, then describe representation extraction, relative measurements and ensemble detection modules.

\subsection{Threat Model and Problem Formulation}
\label{sec:problem_setup}

\paragraph{Threat Model.}
We consider a three-party setting involving a user ($U$), a model service provider ($S$), and a trusted third-party auditor ($D$). 
The provider $S$ deploys a proprietary watermarking algorithm and exposes only a standard text generation API with a binary watermark control flag to the auditor $D$, without revealing the watermarking mechanism, secret keys, or internal model states. 
The user $U$ submits a query text $t_\text{q}$ for verification. 
$D$ is assumed to be a regulated and compliance-certified entity (e.g., an accredited auditing or digital forensics organization) that performs watermark detection independently.
Neither $U$ nor $S$ has access to the detection pipeline, and $D$ has no access to the watermarking algorithm or secret keys. 
This defines a black-box watermark detection problem in which $D$ rely solely on observable output behavior.

\paragraph{Problem Formulation.}
We formalize black-box watermark detection as a hypothesis testing problem. 
Given a query text $t_{\text{q}}$, $D$ evaluates whether $t_{\text{q}}$ is more consistent with $S$'s watermarked output distribution than with its unwatermarked counterpart. 
Since $D$ cannot model the universal distribution of unwatermarked text, we adopt $S$’s unwatermarked output as a local reference anchor for the null hypothesis. 
Specifically, we test:
\begin{equation}
\begin{aligned}
\mathcal{H}_0: t_{\text{q}} \sim P_{\text{o}}, \quad\mathcal{H}_1: t_{\text{q}} \sim P_{\text{wm}},
\end{aligned}
\end{equation}
where $P_{\text{o}}$ and $P_{\text{wm}}$ denote $S$’s unwatermarked and watermarked output distributions, respectively.

\paragraph{TTP-Detect Framework Overview.}
To operationalize the above hypothesis test without access to likelihood functions, $D$ queries $S$ to construct two empirical \textit{reference sets}, $\mathcal{T}_{\text{o}} \sim P_{\text{o}}$ and $\mathcal{T}_{\text{wm}} \sim P_{\text{wm}}$.
Given a query text $t_{\text{q}}$, TTP-Detect first maps the query and reference texts into a proxy-induced representation space to amplify watermark-relevant discrepancies (Sec.~\ref{sec:feature_extraction}). It then applies $M$ complementary \textit{relative measurement modules} to produce scores $\{A_m\}_{m=1}^{M}$, where each $A_m$ measures the affinity of $t_{\text{q}}$ to $\mathcal{T}_{\text{wm}}$ relative to $\mathcal{T}_{\text{o}}$ (Sec.~\ref{sec:local_knn}). Finally, it aggregates and calibrates these signals into a single statistic $\mathrm{Agg}(A_1,\ldots,A_M)$ for decision-making (Sec.~\ref{sec:ensemble}).
Accordingly, unwatermarked texts are expected to yield consistently low relative scores across modules, resulting in an aggregate statistic consistent with $\mathcal{H}_0$. We provide further discussion on practical deployment in Appendix~\ref{sec:appendix_deploy}.

\subsection{Proxy-Based Representation Extraction}
\label{sec:feature_extraction}
Watermark signals are subtle and distributed, making direct detection from raw text difficult in black-box settings.
We therefore extract representations with a \textit{proxy model} adapted to amplify watermark-relevant differences, supported by prior evidence on \textit{watermark learnability}~\cite{gu2024on}.

\paragraph{Discriminative Instruction Tuning.}
To obtain representations that separate watermarked and unwatermarked outputs, we adapt a proxy model via a supervised discrimination objective.
We construct a training set $\mathcal{D}_{\text{sft}}$ by sampling prompts from a corpus and querying $S$ to produce paired completions, each consisting of an unwatermarked output $t_{\text{o}}$ and its watermarked counterpart $t_{\text{wm}}$.
The proxy model $\mathcal{M}_{\text{proxy}}$ is then tuned to predict the corresponding watermark label sequence from the completion text, optimized via conditional negative log-likelihood:
\begin{equation}
\mathcal{L}(\theta) = - \frac{1}{|\mathcal{D}_{\text{sft}}|} \sum_{(t, y) \in \mathcal{D}_{\text{sft}}} \sum_{i=1}^{|y|} \log P_\theta(y_i \vert t, y_{<i}).
\end{equation}

\paragraph{Representation Extraction.}
At detection time, $D$ queries $S$ to build $N$ paired reference samples, yielding two reference sets
$\mathcal{T}_{\text{o}}=\{t_{\text{o}}^{(j)}\}_{j=1}^{N} \sim P_{\text{o}}$ and
$\mathcal{T}_{\text{wm}}=\{t_{\text{wm}}^{(j)}\}_{j=1}^{N} \sim P_{\text{wm}}$.
Given any text $t$ (either the query $t_{\text{q}}$ or a reference sample), we encode it with $\mathcal{M}_{\text{proxy}}$ and extract the $\ell_2$-normalized hidden state of the final token from the last layer, denoted as $z(t)\in\mathbb{R}^{d_{\text{proxy}}}$.
This representation captures both contextual semantics and watermark-discriminative cues internalized during proxy training.
Accordingly, we obtain feature sets $\mathcal{Z}_{\text{o}}=\{z(t_{\text{o}}^{(j)})\}_{j=1}^{N}$ and $\mathcal{Z}_{\text{wm}}=\{z(t_{\text{wm}}^{(j)})\}_{j=1}^{N}$, and represent the query as $z_{\text{q}}=z(t_{\text{q}})$.

\subsection{Relative Measurement Modules}
\label{sec:local_knn}
Watermarking induce bias that does not concentrate on a single statistic, but manifests across different statistical scales of the generated text. As a result, relying on a single detection criterion is brittle under algorithmic diversity and semantic variation. TTP-Detect employs multiple relative measurements, each probing watermark consistency from a complementary statistical perspective.
\subsubsection{Local Consistency Test}
\label{sec:local_density}
Local consistency measures whether a query representation is locally surrounded by watermarked samples, reflecting the tendency of watermark-induced bias to preserve neighborhood-level coherence under similar semantics.

We estimate the local watermark likelihood by examining the neighborhood of the query representation $z_{\text{q}}$ in feature space. 
Let $\mathcal{X}_{\text{ref}} = \mathcal{Z}_{\text{o}} \cup \mathcal{Z}_{\text{wm}}$ denote the combined reference set. 
Since all representations are $\ell_2$-normalized, semantic affinity is measured by cosine similarity. 
We define the local support region $\mathcal{N}(z_{\text{q}})$ as the $k$ nearest neighbors of $z_{\text{q}}$ in $\mathcal{X}_{\text{ref}}$.
To quantify local affinity, we employ a kernel-based density estimator over $\mathcal{N}(z_{\text{q}})$. 
Each neighbor $z_i \in \mathcal{N}(z_{\text{q}})$ contributes a weight
\begin{equation}
    w_i = \exp\left(-\frac{1 - \text{sim}(z_{\text{q}}, z_i)}{\sigma^2}\right),
\end{equation}
where $\sigma$ is an adaptive bandwidth set to the average cosine distance within $\mathcal{N}(z_{\text{q}})$, normalizing for variations in local sample density.

The \textit{local consistency score} is then defined as the normalized density mass of watermarked samples:
\begin{equation}
A_{\text{Loc}}(z_{\text{q}})=
\frac{\sum_{z_i \in \mathcal{N}(z_{\text{q}})\cap \mathcal{Z}_{\text{wm}}} w_i}
     {\sum_{z_i \in \mathcal{N}(z_{\text{q}})} w_i}.
\end{equation}
A higher $A_{\text{Loc}}$ indicates that $z_{\text{q}}$ lies in a neighborhood dominated by watermarked references, providing evidence of local watermark consistency.

\subsubsection{Global Geometry Test}
\label{sec:global_geometry}
While local measurements identify watermark traces on local manifolds, it may fail to capture the systematic distributional shifts. Watermarking operates as a global intervention, often resulting in distinguishable geometric signatures across the entire feature space. We measure this global affinity using two complementary distance-based tests.

\paragraph{Mahalanobis Score.}

To account for correlated watermark perturbations, we evaluate the query using a Mahalanobis distance. 
Given the high dimensionality of representations, we project reference samples into a lower-dimensional subspace via PCA (details in Appendix~\ref{app:geometry_details}) and estimate class-specific regularized covariances. 
The \textit{Mahalanobis score} is defined as
\begin{equation}
    \Delta_{\text{Mah}}(z_{\text{q}}) = \delta_{\hat{\Sigma}_{\text{o}}}^2(\tilde{z}_{\text{q}}, \tilde{\mu}_{\text{o}}) - \delta_{\hat{\Sigma}_{\text{wm}}}^2(\tilde{z}_{\text{q}}, \tilde{\mu}_{\text{wm}}),
\end{equation}
where $\delta_{\Sigma}^2$ denotes the squared Mahalanobis distance in the subspace. A positive $\Delta_{\text{Mah}}$ indicates closer alignment with the watermarked distribution.

\paragraph{Energy Score.}
Some complex watermarking schemes induce irregular distributions that violate the Gaussian assumption of the Mahalanobis test. To handle this, we additionally employ a non-parametric Energy distance.
We define the \textit{Energy score} as the contrast between the energy distances to the two reference sets:
\begin{equation}
    \Delta_{\text{Ene}}(z_{\text{q}}) = \delta_{\text{Ene}}(z_{\text{q}}, \mathcal{Z}_{\text{o}}) - \delta_{\text{Ene}}(z_{\text{q}}, \mathcal{Z}_{\text{wm}}),
\end{equation}
where $\delta_{\text{Ene}}$ measures the average pairwise distance adjusted by the internal dispersion of each reference set (see Appendix~\ref{app:geometry_details} for details).

Both statistics are mapped to normalized scores via a sigmoid function, yielding $A_{\text{Mah}}$ and $A_{\text{Ene}}$.

\subsubsection{Adaptive Rank Test}
\label{sec:ppl_dynamics}
This test targets watermark signals manifested in the generation dynamics rather than in embedding geometry.
Since some schemes perturb token selection probabilities with minimal semantic or representational shift, watermark traces can surface in token-level likelihood patterns even when representation-based measurements are inconclusive.
We therefore compute negative log-likelihood (NLL) based statistics under $\mathcal{M}_{\text{score}}$ and apply a direction-adaptive rank test against paired reference sets.

Given a text $t{=}(x_1,\dots,x_L)$, we compute the token-wise NLL sequence $\boldsymbol{\ell} {=} \{\ell_i\}_{i=1}^L$, where $\ell_i {=} -\log P(x_i \vert x_{<i})$. We then extract two complementary statistics: (1) global cross-entropy $E_{\text{GE}}{=}\frac{1}{L}\sum_{i=1}^{L}\ell_i$, capturing overall fluency; (2) local volatility $E_{\text{LV}}{=}\sqrt{\frac{1}{L}\sum_{i=1}^{L}\big(\ell_i-E_{\text{GE}}\big)^2}$, capturing generation confidence variability.

For a feature $f\in\{E_{\text{GE}},E_{\text{LV}}\}$, the effect direction varies across watermarking schemes.
To avoid assuming a fixed direction, we infer an adaptive direction from the reference sets $\mathcal{T}_{\text{o}}$ and $\mathcal{T}_{\text{wm}}$:
$\rho_f=\operatorname{sign}\!\left(\mathbb{E}_{t\in\mathcal{T}_{\text{wm}}}[f(t)]-\mathbb{E}_{t\in\mathcal{T}_{\text{o}}}[f(t)]\right)$, 
so that larger $\tilde f(t)=\rho_f\cdot f(t)$ is aligned with the watermarked tendency for this feature.

We estimate the conformity ranks of $t_{\text{q}}$ under both reference sets:
\begin{equation}
\begin{aligned}
\footnotesize
p_{\text{wm}}^f
&=\frac{1+\sum_{t\in\mathcal{T}_{\text{wm}}}\mathbb{I}\!\left[\tilde f(t)\le \tilde f(t_{\text{q}})\right]}{|\mathcal{T}_{\text{wm}}|+1},\\
p_{\text{o}}^f
&=\frac{1+\sum_{t\in\mathcal{T}_{\text{o}}}\mathbb{I}\!\left[\tilde f(t)\ge \tilde f(t_{\text{q}})\right]}{|\mathcal{T}_{\text{o}}|+1}.
\end{aligned}
\end{equation}
After alignment, a larger $\tilde f(t_{\text{q}})$ implies larger $p_{\text{wm}}^f$ and smaller $p_{\text{o}}^f$.
We normalize them into a watermark tendency score ($\epsilon$ for numerical stability):
$s_f(t_{\text{q}}){=}{p_{\text{wm}}^f}/({p_{\text{wm}}^f+p_{\text{o}}^f+\epsilon})\in[0,1]$. 

Finally, we fuse global fluency and local volatility into an \textit{adaptive rank score}, where $\lambda$ trades off the overall likelihood shift and local volatility:
\begin{equation}
A_{\text{Ada}}(t_{\text{q}})=\lambda\, s_{E_{\text{GE}}}(t_{\text{q}}) + (1-\lambda)\, s_{E_{\text{LV}}}(t_{\text{q}}).
\end{equation}

\subsection{Ensemble and Detection}
\label{sec:ensemble}
Each relative measurement module captures a complementary aspect of the watermark signal, including local consistency ($A_{\text{Loc}}$), global geometry ($A_{\text{Mah}}, A_{\text{Ene}}$), and adaptive rank ($A_{\text{Ada}}$). Given the diversity of watermarking mechanisms, no single score is universally reliable. We therefore combine these measurements through a lightweight ensemble.
For a query text $t_{\text{q}}$, we form a score vector $\mathbf{A}(t_{\text{q}}) {=} [A_{\text{Loc}}, A_{\text{Mah}}, A_{\text{Ene}}, A_{\text{Ada}}]^\top \in [0,1]^4$ and compute the \textit{ensemble score} as
\begin{equation}
    A_{\text{ens}}(t_{\text{q}}) = \sigma\left( \mathbf{w}^\top \mathbf{A}(t_{\text{q}}) + b \right),
\end{equation}
where $\mathbf{w} \in \mathbb{R}^4$ and $b \in \mathbb{R}$ are learnable parameters, and $\sigma$ is the sigmoid function.

\paragraph{Robust Calibration Strategy.}
To estimate appropriate weights that remain effective even under potential evasion attacks, we construct an augmented validation set $\mathcal{D}_{\text{val}}$ that includes both standard reference pairs and their adversarially perturbed counterparts (e.g., via paraphrasing or editing). We model the ensemble weights using logistic regression by minimizing the binary cross-entropy loss on $\mathcal{D}_{\text{val}}$.

\paragraph{Decision Rule.}
To support third-party auditing, the final decision threshold $\tau$ is selected to explicitly control the false positive rate (FPR). We calibrate $\tau$ on a large held-out set of benign, unwatermarked texts:
\begin{equation}
    \tau = \inf \left\{ \gamma \in (0,1) : \widehat{\text{FPR}}(\gamma; \mathcal{D}_{\text{benign}}) \le \alpha \right\},
\end{equation}
where $\alpha$ is the target FPR. The final verdict is given by $ \hat{y} = \mathbb{I}(A_{\text{ens}}(t_{\text{q}}) \ge \tau)$, where $\widehat{\text{FPR}}$ denotes the empirical false positive rate. The large sample size of $\mathcal{D}_{\text{benign}}$ ensures that the estimated threshold is statistically significant and reliable. 

TTP-Detect offers significant extensibility. As new detection techniques emerge, they can be seamlessly integrated as additional feature channels in $\mathbf{A}$ without altering the core architecture.
\section{Experiments}
\begin{table*}[t]
\centering
\scriptsize
\setlength{\tabcolsep}{4.7pt}             
\begin{tabular}{ccccccccccccccccc}  
\toprule
      & \multicolumn{8}{c}{\textsc{C4 Dataset}}
      & \multicolumn{8}{c}{\textsc{OpenGen Dataset}} \\
\cmidrule(lr){2-9} \cmidrule(lr){10-17}
\textbf{Watermark}
    & \multicolumn{4}{c}{Llama-3.1-8B} 
    & \multicolumn{4}{c}{OPT-6.7B} 
    & \multicolumn{4}{c}{Llama-3.1-8B} 
    & \multicolumn{4}{c}{OPT-6.7B}\\
\cmidrule(lr){2-5} \cmidrule(lr){6-9}
\cmidrule(lr){10-13} \cmidrule(lr){14-17}
      & TPR & TNR & F1 & AUC
      & TPR & TNR & F1 & AUC
      & TPR & TNR & F1 & AUC
      & TPR & TNR & F1 & AUC \\
\specialrule{\lightrulewidth}{\aboverulesep}{1pt}
\addlinespace[1pt]
UPV & 0.985 & 0.980 & 0.983 & 0.991 & 0.990 & 0.990 & 0.990 & 0.998 & 0.995 & 0.960 & 0.978 & 0.994 & 0.995 & 0.980 & 0.988 & 0.996 \\ 
\specialrule{\lightrulewidth}{\aboverulesep}{0pt}
\rowcolor{gray!15}\multicolumn{17}{c}{\textit{Logits-based Schemes}} \\
\addlinespace[1pt]
$\text{KGW}_\texttt{TTP}$ & 0.980 & 0.980 & 0.980 & 0.998 & 1.000 & 0.990 & 0.995 & 0.999 & 0.980 & 0.950 & 0.966 & 0.993 & 0.980 & 1.000 & 0.990 & 0.999  \\
$\text{Unigram}_\texttt{TTP}$ & 1.000 & 0.990 & 0.995 & 0.999 & 1.000 & 1.000 & 1.000 & 1.000 & 1.000 & 1.000 & 1.000 & 1.000 & 1.000 & 1.000 & 1.000 & 1.000  \\
$\text{SWEET}_\texttt{TTP}$ & 0.985 & 0.965 & 0.975 & 0.997 & 0.995 & 0.995 & 0.995 & 0.999 & 0.980 & 0.920 & 0.951 & 0.981 & 0.980 & 0.970 & 0.975 & 0.997  \\
$\text{MorphMark}_\texttt{TTP}$ & 0.945 & 0.965 & 0.955 & 0.981 & 0.990 & 0.940 & 0.966 & 0.993 & 0.925 & 0.835 & 0.885 & 0.942 & 0.925 & 0.965 & 0.944 & 0.989  \\
\rowcolor{gray!15}\multicolumn{17}{c}{\textit{Distribution-preserving Schemes}} \\
\addlinespace[1pt]
$\text{Unbiased}_\texttt{TTP}$ & 0.870 & 0.845 & 0.859 & 0.911 & 0.865 & 0.870 & 0.867 & 0.920 & 0.795 & 0.775 & 0.787 & 0.838 & 0.775 & 0.760 & 0.769 & 0.826  \\
$\text{SynthID}_\texttt{TTP}$ & 0.865 & 0.930 & 0.894 & 0.938 & 0.910 & 0.905 & 0.908 & 0.957 & 0.875 & 0.785 & 0.838 & 0.896 & 0.860 & 0.855 & 0.858 & 0.924  \\
\rowcolor{gray!15}\multicolumn{17}{c}{\textit{Synthetic Scheme}} \\
\addlinespace[1pt]
$\text{SymMark}_\texttt{TTP}$ & 1.000 & 1.000 & 1.000 & 1.000 & 1.000 & 1.000 & 1.000 & 1.000 & 1.000 & 1.000 & 1.000 & 1.000 & 1.000 & 1.000 & 1.000 & 1.000  \\
\bottomrule
\end{tabular}
\caption{Watermark detection performance of various watermarking schemes under TTP-Detect framework.}
\label{tab:detectability}
\end{table*}

\subsection{Experimental Setup}
\paragraph{Models.}
We employ three families of LLMs to instantiate the different roles in our framework. On the service provider side, we use Llama-3.1-8B~\cite{grattafiori2024llama3herdmodels} and OPT-6.7B \cite{zhang2022optopenpretrainedtransformer} as generation models. On the detector (TTP) side, we use Qwen2.5-3B~\cite{qwen2025qwen25technicalreport} as the proxy model for representation extraction, and use Qwen3-1.7B~\cite{yang2025qwen3technicalreport} as the scoring model in the adaptive rank test module.

\paragraph{Watermarking Schemes.}
We choose UPV~\cite{liu2024an}, which supports detection without access to a secret key, as the black-box watermark detection baseline. Under the TTP-Detect framework, we perform black-box detection for seven watermarking schemes, including KGW~\cite{pmlr-v202-kirchenbauer23a}, Unigram~\cite{zhao2024provable}, SynthID~\cite{dathathri2024scalable}, SWEET~\cite{lee-etal-2024-wrote}, Unbiased~\cite{hu2024unbiased}, SymMark~\cite{wang-etal-2025-trade}, and MorphMark~\cite{wang-etal-2025-morphmark}. Details are provided in Appendix~\ref{sec:appendix_watermark-schemes}.

\paragraph{Datasets and Evaluation Metrics.}
To stay aligned with standard evaluation practices in LLM watermark detection, we adopt the widely used news-like C4~\cite{10.5555/3455716.3455856} and long-form OpenGen~\cite{krishna2023paraphrasing} datasets to assess detection performance under both in-domain and distribution-shifted settings. We report TPR, TNR, best F1, and AUROC as the primary metrics. For robustness under different attack settings, we additionally plot ROC curves, which characterize the trade-off between FPR and TPR across varying decision thresholds.

\paragraph{Implementation Details.}
We treat each pair of $(\text{watermarking scheme}, \text{generation model})$ as an independent setting. To train the proxy model, we construct an instruction-tuning corpus from C4 by sampling 6{,}000 segments, taking the first 30 tokens as prompts, and generating matched watermarked and unwatermarked continuations that are then wrapped into an instruction template. Separately, we use additional C4-held-out data to tune module-specific hyperparameters and to estimate the ensemble weights from per-module scores. All reported numbers are computed on disjoint evaluation data, with OpenGen used exclusively to assess out-of-domain generalization. During detection, we use the first 50 tokens of the queried text as a prompt and sample $N{=}16$ reference texts under both watermarked and unwatermarked conditions. Further implementation details are provided in Appendix~\ref{sec:appendix_implementation-details}, and hyperparameter analysis is provided in Appendix~\ref{sec:appendix_hyperparameter}.

\subsection{Main Results}
Table~\ref{tab:detectability} shows that watermark detection performance under the TTP-Detect framework varies across watermark families. Specifically, although we include UPV as a baseline, its detector network is trained for a particular generator network, which is more lenient than TTP-Detect's settings. 

For classic logits-based schemes, $\text{KGW}_\texttt{TTP}$ achieves F1 and AUC on both C4 and OpenGen that exceed or closely match UPV, while $\text{Unigram}_\texttt{TTP}$ consistently attains AUC no lower than 0.999. These results indicate that the traditional red-green list paradigm induces a clear separation between watermarked and unwatermarked texts, and such signals can be effectively captured by TTP-Detect.  
For improved schemes SWEET and MorphMark, although they incorporate dynamic thresholding and adaptive watermark strength, they remain fundamentally rooted in the red-green list paradigm and therefore still exhibit salient, detectable signals. Notably, $\text{SWEET}_\texttt{TTP}$ incurs only a 0.38\% average AUC drop compared to $\text{KGW}_\texttt{TTP}$. 

For the synthetic watermark SymMark, TTP-Detect reaches perfect performance (F1 and AUC of \textbf{1.000}) on all datasets and models, demonstrating strong perceptibility to its synthesized watermark patterns.
For distribution-preserving methods such as SynthID and Unbiased, we evaluate under their original non-distortion setting. Compared to biased watermarking, the induced distributional differences between watermarked and unwatermarked texts are substantially weaker, making detection more challenging; nonetheless, TTP-Detect is still able to capture measurable discrepancies.

\subsection{Robustness to Attacks}
\begin{figure}[t]
  \centering
  \includegraphics[width=\linewidth]{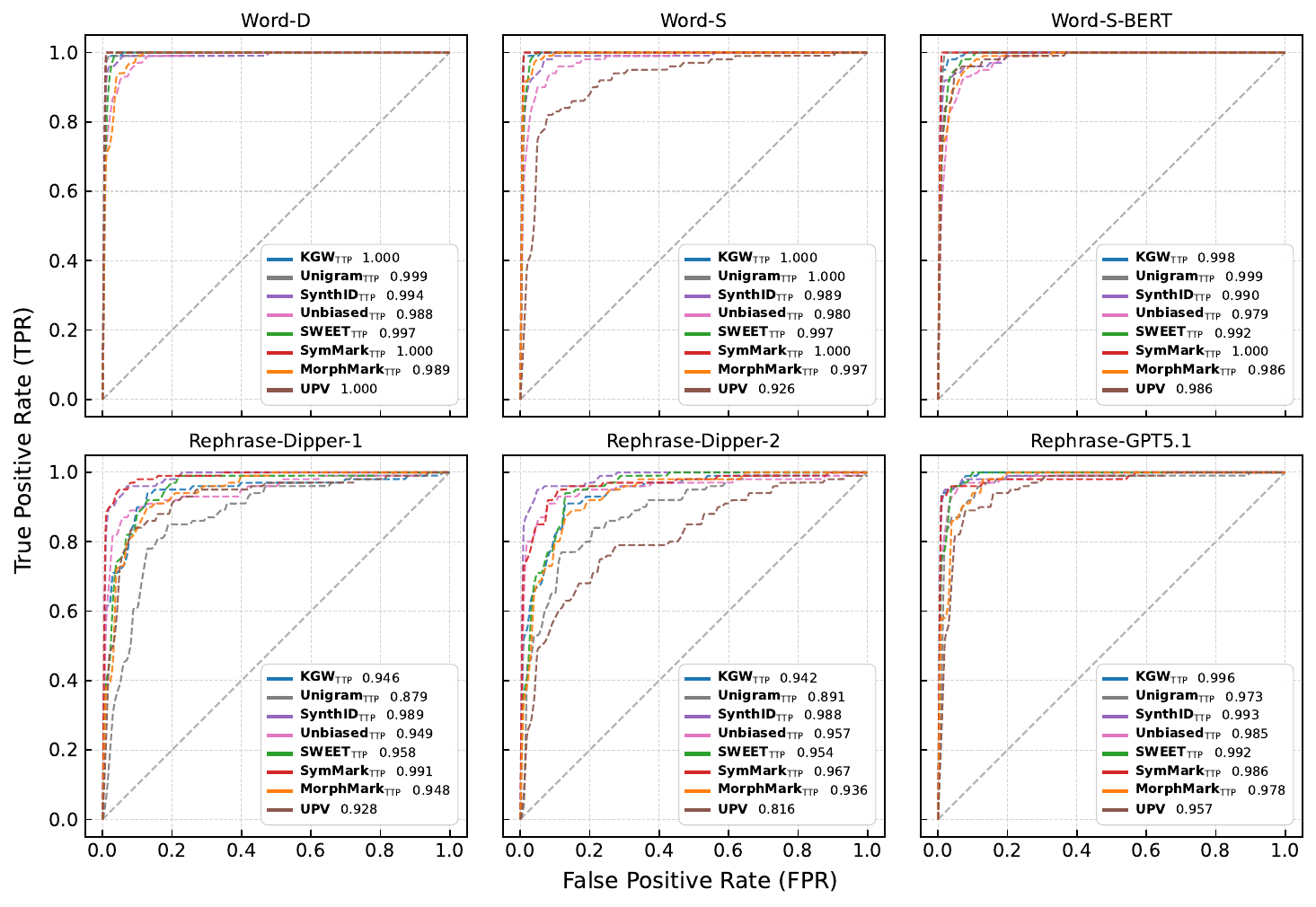}
  \caption{Robustness across various attack types.}
  \label{fig:robustness_roc} 
\end{figure}
To evaluate the robustness of TTP-Detect, we attack watermarked texts generated by Llama-3.1-8B on C4 with six transformations grouped into two categories: \emph{editing}-based attacks (Word-D, Word-S, Word-S-BERT) and \emph{paraphrasing}-based attacks (Dipper-1, Dipper-2, GPT-5.1). Figure~\ref{fig:robustness_roc} reports the main results, while Appendix~\ref{sec:appendix_robustness} provides the full attack configurations.
The ROC curves and AUC scores show that TTP-Detect remains highly robust across a broad range of attacks, e.g., achieving an average AUC of 0.980 for KGW$_{\texttt{TTP}}$. This robustness stems from the framework’s relative measurement formulation and its integration of multiple complementary modules, which reduces reliance on any single feature or failure mode. Among all evaluated perturbations, Dipper-1 and Dipper-2 are the most disruptive. In particular, Unigram$_{\texttt{TTP}}$ experiences a more noticeable degradation, which we attribute to the proxy model’s strong specialization to Unigram’s fixed vocabulary partitioning; consequently, proxy-dependent modules become less reliable under paraphrases that introduce substantial lexical substitutions and sentence-level reorganization. Nevertheless, under most attack types and for the majority of watermarking schemes, TTP-Detect tends to outperform UPV, suggesting improved robustness under realistic post-generation edits.

\subsection{Ablation Analysis}

\begin{figure}[t]
    \centering
    \begin{subfigure}[t]{0.5\columnwidth}
        \centering
        \includegraphics[width=\linewidth]{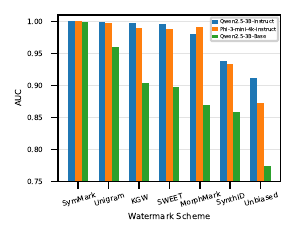}
        \caption{Proxy model ablation}
        \label{fig:ablation_proxy}
    \end{subfigure}\hfill
    \begin{subfigure}[t]{0.5\columnwidth}
        \centering
        \includegraphics[width=\linewidth]{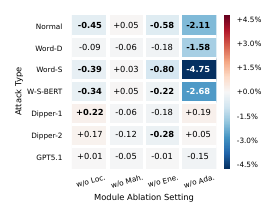}
        \caption{Module ablation}
        \label{fig:ablation_module}
    \end{subfigure}
    \caption{Ablation study for (a) proxy model and (b) relative measurement modules.}
    \label{fig:ablation}
\end{figure}

\paragraph{Impact of Proxy Model Choice.}
In our main experiments, we use LoRA-tuned Qwen2.5-3B-Instruct as the proxy model. To assess the sensitivity of this choice, we replace it with a comparably sized model from a different family, Phi-3-mini-4k-instruct~\cite{abdin2024phi3technicalreporthighly}, while keeping the rest of the pipeline unchanged. As shown in Figure~\ref{fig:ablation_proxy}, the two proxies yield very similar performance across watermarking schemes, with the average AUC differing by less than 1\%, suggesting that TTP-Detect is not tied to any particular model family.
We further test whether proxy fine-tuning is essential by using an untuned Qwen2.5-3B-Base model as the proxy, where we extract text representations by mean-pooling the last-layer hidden states. Although performance degrades compared to instruction-tuned proxies, TTP-Detect remains effective without proxy training. In particular, under SymMark$_{\texttt{TTP}}$ and Unigram$_{\texttt{TTP}}$, the AUC drop relative to the instruction-tuned proxy is below 4\%, indicating that proxy fine-tuning primarily amplifies watermark-sensitive cues, rather than being a prerequisite for the relative measurement.

\paragraph{Impact of Relative Measurement Modules.}
We conduct a leave-one-out ablation: for each watermarking scheme, we remove one of the four modules and measure the change in AUC under every attack scenario, while keeping all other settings fixed. Figure~\ref{fig:ablation_module} visualizes the ablation results for SynthID$_{\texttt{TTP}}$, where each cell denotes the AUC change relative to the full system. The Adaptive Rank Test module is the most influential component: removing it yields the largest AUC drops in most scenarios, even though it can slightly improve performance under the two Dipper paraphrase settings. This pattern highlights the intended complementarity of our design: a module may be particularly valuable for broad robustness even if it is not uniformly beneficial under every specific attack. We provide the corresponding analyses for the remaining watermarking schemes in Appendix~\ref{sec:module-ablation}.

\subsection{Further Analysis}
\label{sec:further-analysis}
\paragraph{Impact of the Size of Reference Set.}

\begin{figure}[t]
    \centering
    \begin{subfigure}[t]{0.49\columnwidth}
        \centering
        \includegraphics[width=\linewidth]{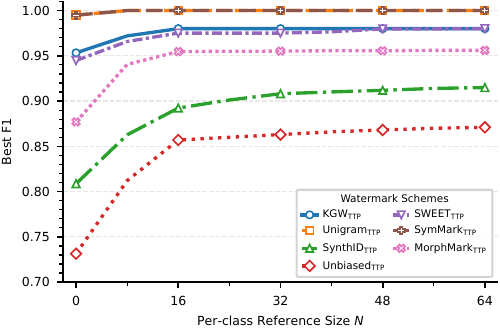}
        \caption{Size of reference set}
        \label{fig:reference_size}
    \end{subfigure}\hfill
    \begin{subfigure}[t]{0.49\columnwidth}
        \centering
        \includegraphics[width=\linewidth]{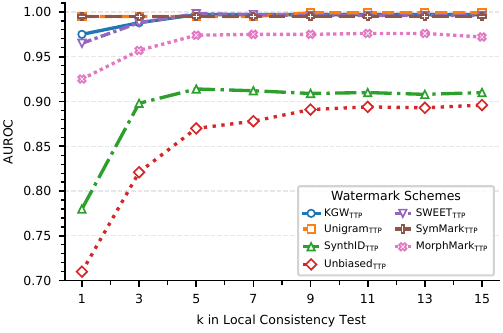}
        \caption{Hyperparameter $k$}
        \label{fig:hyperparameter_k}
    \end{subfigure}
    \caption{(a) Impact of the size of reference set; (b) hyperparameter analysis of $k$ in Local Consistency Test.}
    \label{fig:further_analysis}
\end{figure}
We vary the per-class reference set size as $N\in\{0,8,16,32,48,64\}$ and report the resulting best F1 in Figure~\ref{fig:reference_size}. Here, $N{=}0$ corresponds to a reference-free setting, where we directly rely on the tuned proxy to classify the query based on its output without relative comparisons. Overall, incorporating a reference set consistently improves performance across all watermark schemes. For schemes with stronger signals (e.g., Unigram and SymMark), the proxy already exhibits substantial separability, and the gains from increasing $N$ are modest. In contrast, for distribution-preserving schemes (e.g., Unbiased and SynthID), using $N{=}8$ and $N{=}16$ yields clear benefits over $N{=}0$, improving average best F1 by more than 5\% and 10\%, respectively. Beyond $N{=}16$, the improvements taper off, indicating diminishing returns. We therefore adopt $N{=}16$ as a practical trade-off between effectiveness and cost.

\paragraph{Efficiency Analysis.}
\begin{table}
  \centering
  \scriptsize
  \begin{tabular}{llc}
    \hline
    \textbf{Phase/Module} & \textbf{Model for Test} & \textbf{Time (s)} \\
    \hline
    \rowcolor{gray!15} 
    \multicolumn{3}{c}{\textit{Service Provider Side}} \\
    Generation of Reference Set & Llama-3.1-8B & 6.1374\\
    \rowcolor{gray!15} 
    \multicolumn{3}{c}{\textit{TTP Side}} \\
    Representation Extraction & Qwen2.5-3B & 0.8348\\
    Local Consistency Test & $-$ & 0.0012\\
    Relative Mahalanobis Score & $-$ & 0.0165\\
    Relative Energy Score & $-$ & 0.0113 \\
    Adaptive Rank Test & Qwen3-1.7B & 1.7328\\
    \hline
  \end{tabular}
  \caption{Efficiency Analysis.}
  \label{tab:efficiency}
\end{table}
Table~\ref{tab:efficiency} reports the efficiency results under a reference set size of $N=16$ and a generation length of 300 tokens. The dominant cost of TTP-Detect comes from invoking the service provider to generate the reference samples. With batched inference on Llama-3.1-8B, this step completes in 6.14s. On the TTP side, modules can be executed in parallel. The modules that rely on proxy-model representation extraction (Local Consistency Test and the Mahalanobis/Energy scores) finish within 1s in total, while the Adaptive Rank Test completes in 1.74s. Overall, the full TTP-side computation stays below 2s.

\paragraph{Hyperparameter Analysis.} 
Figure~\ref{fig:hyperparameter_k} analyzes the hyperparameter $k$ in the Local Consistency Test module; additional details and other hyperparameters are discussed in Appendix~\ref{sec:appendix_hyperparameter}.

\section{Conclusion}
This paper presents TTP-Detect, a non-intrusive black-box verification framework that decouples detection from watermark design. By casting verification as a relative hypothesis test using paired watermarked/unwatermarked reference samples, it avoids reliance on private keys and scheme-specific detectors. Experiments demonstrate strong detectability and robustness under realistic attacks across representative watermark families and generation models. These gains stem from a proxy representation space and complementary relative measurements, combined via a lightweight calibrated ensemble. Efficiency results show modest TTP-side overhead, supporting scalable third-party watermark audits.

\section*{Limitations}

TTP-Detect decouples watermark verification from injection, providing a practical pathway toward neutral, third-party auditing in black-box settings. While the framework is agnostic to the watermarking algorithm, underlying model and secret key, its gains can be less pronounced for schemes explicitly engineered to minimize bias or distortion. This is because TTP-Detect’s relative measurement modules primarily exploit observable distributional and generative discrepancies between watermarked and non-watermarked outputs; when such discrepancies are intentionally suppressed, the available signal becomes inherently weaker. A promising direction is to broaden the measurement suite with additional relative tests that capture subtler watermark-induced effects. Nonetheless, TTP-Detect constitutes an important step toward deployable black-box watermark verification as an auditing primitive, helping bridge the gap between watermark design and real-world accountability.


\bibliography{custom}
\clearpage
\appendix

\section{Details of Global Geometry Test}
\label{app:geometry_details}

\paragraph{Notation.}
Let $\mathcal{Z}_{\text{o}}=\{z^i_{\text{o}}\}_{i=1}^{N}$ and $\mathcal{Z}_{\text{wm}}=\{z^i_{\text{wm}}\}_{i=1}^{N}$ denote the $\ell_2$-normalized representations of the non-watermarked and watermarked reference sets, respectively.
We write the joint reference set as $\mathcal{X}_{\text{ref}}=\mathcal{Z}_{\text{o}}\cup\mathcal{Z}_{\text{wm}}$ with $|\mathcal{X}_{\text{ref}}|=2N$.

\paragraph{PCA Projection.}
To mitigate covariance singularity and reduce noise in high-dimensional spaces, we fit PCA on the centered joint reference set.
Let the empirical mean be
\begin{equation}
    \mu_{\text{ref}} = \frac{1}{2N}\sum_{z\in\mathcal{X}_{\text{ref}}} z .
\end{equation}
Let $W\in\mathbb{R}^{d\times d'}$ be the matrix of the top-$d'$ principal directions computed from
$\{z-\mu_{\text{ref}}: z\in\mathcal{X}_{\text{ref}}\}$.
In practice $d'$ is set to a small value satisfying $d'\le \min(d,2N-1)$ to respect the rank constraint after centering.
For any representation $z$, we define its projected vector as
\begin{equation}
    \tilde{z} = W^\top (z-\mu_{\text{ref}})\in\mathbb{R}^{d'} .
\end{equation}
We then compute class means in the subspace:
\begin{equation}
    \tilde{\mu}_c = \frac{1}{N}\sum_{z\in \mathcal{Z}_c}\tilde{z},\quad c\in\{\text{o},\text{wm}\}.
\end{equation}

\paragraph{Regularized Mahalanobis Distance.}
Let $\tilde{\Sigma}_c$ be the empirical covariance of projected vectors in class $c$:
\begin{equation}
    \tilde{\Sigma}_c = \frac{1}{N-1}\sum_{z\in\mathcal{Z}_c} (\tilde{z}-\tilde{\mu}_c)(\tilde{z}-\tilde{\mu}_c)^\top .
\end{equation}
We apply shrinkage regularization to ensure invertibility:
\begin{equation}
    \hat{\Sigma}_c = \tilde{\Sigma}_c + \alpha \cdot \frac{\mathrm{tr}(\tilde{\Sigma}_c)}{d'} \mathbf{I}_{d'},
\end{equation}
where $\alpha$ is a hyperparameter and $\mathbf{I}_{d'}$ is the identity matrix.
The squared Mahalanobis distance used in the main text is
\begin{equation}
    \delta_{\hat{\Sigma}_c}^2(\tilde{z},\tilde{\mu}_c)
    =(\tilde{z}-\tilde{\mu}_c)^\top \hat{\Sigma}_c^{-1}(\tilde{z}-\tilde{\mu}_c).
\end{equation}
Accordingly, the Mahalanobis contrast in the main text is
\begin{equation}
    \Delta_{\text{Mah}}(z_{\text{q}})
    = \delta_{\hat{\Sigma}_{\text{o}}}^2(\tilde{z}_{\text{q}},\tilde{\mu}_{\text{o}})
    - \delta_{\hat{\Sigma}_{\text{wm}}}^2(\tilde{z}_{\text{q}},\tilde{\mu}_{\text{wm}}).
\end{equation}

\paragraph{Energy Distance Definition.}
To avoid parametric assumptions, we additionally compute a non-parametric energy statistic.
We use the following query-to-set energy distance:
\begin{equation}
\footnotesize
    \delta_{\text{Ene}}(z_{\text{q}}, \mathcal{Z}_c)
    = \frac{2}{N}\sum_{z\in\mathcal{Z}_c}\|z_{\text{q}}-z\|_2
      -\frac{1}{N^2}\sum_{z,z'\in\mathcal{Z}_c}\|z-z'\|_2,
\end{equation}
where the first term measures the average potential energy between the query and the set, and the second term subtracts the internal dispersion of the set itself.
The energy contrast in the main text is then
\begin{equation}
    \Delta_{\text{Ene}}(z_{\text{q}})
    = \delta_{\text{Ene}}(z_{\text{q}}, \mathcal{Z}_{\text{o}})
    - \delta_{\text{Ene}}(z_{\text{q}}, \mathcal{Z}_{\text{wm}}).
\end{equation}

\paragraph{Score Normalization.}
As stated in Sec.~\ref{sec:global_geometry}, we map both geometry contrasts to $[0,1]$ using a sigmoid:
\begin{equation}
\begin{aligned}
    A_{\text{Mah}}(z_{\text{q}})&=\sigma\!\big(\beta_{\text{Mah}}\Delta_{\text{Mah}}(z_{\text{q}})\big),\\
    A_{\text{Ene}}(z_{\text{q}})&=\sigma\!\big(\beta_{\text{Ene}}\Delta_{\text{Ene}}(z_{\text{q}})\big),
\end{aligned}
\end{equation}
where $\sigma(x)=\frac{1}{1+\exp(-x)}$.
$\beta_{\text{Mah}}$ and $\beta_{\text{Ene}}$ are temperature (scale) parameters used to avoid saturation and make the two statistics numerically comparable before ensembling.

\section{Details of Watermarking Schemes}
\label{sec:appendix_watermark-schemes}
We take UPV~\cite{liu2024an} as our main comparison baseline and evaluate the black-box detection of seven target watermarking schemes under our TTP-Detect Framework.

\paragraph{Baseline.}
We take UPV~\cite{liu2024an} as our main comparison baseline. Since the original paper does not release pre-trained watermark generation/detection networks for our selected generation models and datasets, we re-train UPV based on the official implementation. Concretely, we train UPV’s generator and detector networks on C4 and OpenGen for Llama-3.1-8B and OPT-6.7B, respectively. Following UPV’s design, we set the \texttt{bit\_number} to match the corresponding vocabulary size (17 for Llama-3.1-8B and 16 for OPT-6.7B), while keeping the remaining hyperparameters consistent across all settings. 

We also noticed that there are two more works that are related to third-party or publicly verifiable watermarking, namely PVMark~\cite{duan2025pvmarkenablingpublicverifiability} and PDW~\cite{fairoze2025publiclydetectablewatermarkinglanguagemodels}. However, we didn't choose them as baseline for reasons.
\begin{itemize}[leftmargin=*]
    \item PVMark is not a standalone public detector, but a zero-knowledge-proof (ZKP) plugin that makes a keyed watermark detector publicly verifiable: an owner/authorized party runs watermark detection without revealing the secret key, and third parties mainly verify the correctness proof rather than performing key-free detection themselves. Moreover, PVMark requires scheme-specific ZKP-friendly redesign of the original embedding/detection pipeline, which changes the implementation assumptions and is orthogonal to our black-box, parameter-free third-party detection setting. Finally, while PVMark reports implementations across multiple languages, the paper does not provide an officially released open-source repository for reproduction.
    \item PDW proposes a new publicly-detectable watermarking scheme that embeds a publicly verifiable cryptographic signature into model outputs via a rejection-sampling-based decoding procedure. This setting assumes the service provider is willing to (i) modify the generation pipeline to perform signature-conditioned sampling and (ii) publish the public verification key required for detection. As a result, PDW is not a key-free third-party verification baseline for existing private-key watermarks under black-box/API access; rather, it changes the watermarking design and deployment assumptions. Moreover, its signature-driven rejection sampling introduces non-trivial generation overhead and is difficult to align with our cost-aware evaluation protocol that treats the provider as a standard text-generation API.
\end{itemize}

\paragraph{Target Watermarking Schemes.}
We evaluate the black-box detectability of KGW~\cite{pmlr-v202-kirchenbauer23a}, Unigram~\cite{zhao2024provable}, SynthID~\cite{dathathri2024scalable}, SWEET~\cite{lee-etal-2024-wrote}, Unbiased~\cite{hu2024unbiased}, SymMark~\cite{wang-etal-2025-trade} and MorphMark~\cite{wang-etal-2025-morphmark} under the TTP-Detect framework. To align with prior work, we adopt the same hyperparameter settings as the open-source MarkLLM repository~\cite{pan-etal-2024-markllm} for all schemes. For SymMark, we follow the original paper and use UniEXP, a hybrid symbiotic configuration that combines Unigram and AAR~\cite{aaronson2023watermarking}, and we adopt the hyperparameters from the authors’ official implementation. The complete configurations are summarized in Table~\ref{tab:markllm_watermark_parameters}.

\paragraph{}

\begin{table*}[htbp]
    \centering
    \footnotesize
    \begin{tabular}{llll}
        \hline
        \textbf{Algorithm} & \textbf{Source} & \textbf{Parameter} & \textbf{Value} \\
        \hline
        \textbf{UPV} & ~\citet{liu2024an} & \texttt{window\_size} & 3 \\
                                         && \texttt{layers} & 5 \\
                                         && \texttt{use\_sampling} & "True" \\
                                         && \texttt{delta} & 2.0 \\
                                         && \texttt{sampling\_temp} & 0.7 \\
                                         && \texttt{train\_num\_samples} & 10000 \\
                                         && \texttt{bit\_number} & 16(OPT-6.7B) / 17(Llama-3.1-8B) \\
        \hline
        \textbf{KGW} & ~\citet{pmlr-v202-kirchenbauer23a} & $\gamma$ (\texttt{gamma}) & 0.5 \\
                                                         && $\delta$ (\texttt{delta}) & 2.0 \\
                                                         && \texttt{z\_threshold} & 4.0 \\
                                                         && \texttt{prefix\_length} & 1 \\
                                                         && \texttt{f\_scheme} & "time" \\
                                                         && \texttt{window\_scheme} & "left" \\
                                                         && \texttt{hash\_key} & 15485863 \\
        \hline
        \textbf{Unigram} & ~\citet{zhao2024provable} & $\gamma$ (\texttt{gamma}) & 0.5 \\
                                                    && $\delta$ (\texttt{delta}) & 2.0 \\
                                                    && \texttt{z\_threshold} & 4.0 \\
                                                    && \texttt{hash\_key} & 15485863 \\

        \hline
        \textbf{SWEET} & ~\citet{lee-etal-2024-wrote} & $\gamma$ (\texttt{gamma}) & 0.5 \\
                                                    && $\delta$ (\texttt{delta}) & 2.0 \\
                                                    && \texttt{z\_threshold} & 4.0 \\
                                                    && \texttt{hash\_key} & 15485863 \\
                                                    && \texttt{prefix\_length} & 1 \\
                                                    && \texttt{entropy\_threshold} & 0.9 \\
        \hline
        \textbf{SynthID} & ~\citet{dathathri2024scalable} & \texttt{ngram\_len} & 5 \\
                                                         && \texttt{sampling\_table\_size} & 65536 \\
                                                         && \texttt{sampling\_table\_seed} & 0 \\
                                                         && \texttt{watermark\_mode} & "non-distortionary" \\
                                                         && \texttt{num\_leaves} & 2 \\
                                                         && \texttt{context\_history\_size} & "mean" \\
                                                         && \texttt{threshold} & 0.52 \\
        \hline
        \textbf{Unbiased} & ~\citet{hu2024unbiased} & \texttt{type} & "gamma" \\
                                                  && \texttt{n\_grid} & 10 \\
                                                  && \texttt{key} & 42 \\
                                                  && \texttt{prefix\_length} & 5 \\
                                                  && \texttt{p\_threshold} & 0.0005 \\
        \hline
        \textbf{SymMark} & ~\citet{wang-etal-2025-trade} & $\gamma$ (\texttt{gamma}) & 0.25 \\
        (UniEXP)                                       && $\delta$ (\texttt{delta}) & 4.0 \\
                                                       && \texttt{unigram\_hash\_key} & 0 \\
                                                       && \texttt{z\_threshold} & 4.0 \\
                                                       && \texttt{prefix\_length} & 4 \\
                                                       && \texttt{exp\_hash\_key} & 15485863 \\
                                                       && \texttt{threshold} & 1e-4 \\
                                                       && \texttt{top\_k} & 0 \\
                                                       && \texttt{token\_entropy\_threshold} & 2 \\
                                                       && \texttt{semantic\_entropy\_threshold} & 1 \\
                                                       && \texttt{k\_means\_top\_k} & 64 \\
                                                       && \texttt{k\_means\_n\_clusters} & 10 \\
        \hline
        \textbf{MorphMark} & ~\citet{wang-etal-2025-morphmark} & $\gamma$ (\texttt{gamma}) & 0.5 \\
                                                             && \texttt{type} & "exp" \\
                                                             && \texttt{k\_linear} & 1.55 \\
                                                             && \texttt{k\_exp} & 1.3 \\
                                                             && \texttt{k\_log} & 2.15 \\
                                                             && \texttt{p\_0} & 0.15 \\
                                                             && \texttt{hash\_key} & 15485863 \\
                                                             && \texttt{prefix\_length} & 1 \\
                                                             && \texttt{z\_threshold} & 2.0 \\
                                                             && \texttt{f\_scheme} & "time" \\
                                                             && \texttt{window\_scheme} & "left" \\
        \hline
    \end{tabular}
    \caption{Parameter configuration for chosen watermarking algorithms.}
    \label{tab:markllm_watermark_parameters}
\end{table*}

\section{Further Implementation Details}
\label{sec:appendix_implementation-details}
\paragraph{System Configuration.}
All experiments are conducted on a CentOS Linux 7 (Core) server which equipped with dual-socket Intel Xeon Platinum 8375C CPUs (64 physical cores, 128 hardware threads in total) and 2 NVIDIA A100 80GB PCIe GPUs. Our software stack is based on Python~3.12.2, PyTorch~2.5.0 with CUDA~12.1, and Transformers~4.52.4.

\paragraph{Training of Proxy Model for Representation Extraction.}
\begin{figure*}[t]
  \centering
  \begin{tcolorbox}[
      width=\textwidth,
      colback=gray!5,
      colframe=black!70,
      colbacktitle=black!80,
      coltitle=white,
      fonttitle=\bfseries,
      title=Instruction Template for the Training of Representation Extraction Proxy Models,
      rounded corners,
      boxrule=0.6pt
  ]
  {\small\ttfamily
  \textbf{System}: "You are a helpful assistant that helps people detect whether the text is watermarked or not."\\
  \textbf{Instruction}: "Please determine whether the following text is watermarked or not:"\\
  \textbf{Input}: [Text content to be detected...]\\
  \textbf{Response\_positive}: "This text is watermarked."\\
  \textbf{Response\_negative}: "This text is unwatermarked."
  }
  \end{tcolorbox}
  \caption{Instruction template used to construct the LoRA fine-tuning data for representation extraction models.}
  \label{fig:template-for-lora-train}
\end{figure*}
We adopt Qwen2.5-3B-Instruct~\cite{qwen2025qwen25technicalreport} as the representation extractor for main experiments and apply LoRA~\cite{hu2022lora} fine-tuning using the LLaMAFactory~\cite{zheng2024llamafactory} framework. For each (watermarking scheme, generation model) pair, we sample 6,000 segments from C4, use the first 30 tokens as prompts, and query the generator to produce paired watermarked and unwatermarked continuations; we then wrap each completion with an instruction-tuning template to form training instances. The goal is to endow the model with a preliminary ability to distinguish between watermarked and unwatermarked texts, while largely preserving its original semantic understanding capability. The construction of the instruction-tuning dataset is illustrated in Figure~\ref{fig:template-for-lora-train}, and the configuration for fine-tuning is illustrated in Table~\ref{tab:llamafactory-params}.
\begin{table}
    \centering
    \begin{tabular}{ll}
        \hline
        \textbf{Parameter} & \textbf{Value} \\
        \hline
        \texttt{finetuning\_type} & "lora" \\
        \texttt{lora\_rank} & 8 \\
        \texttt{lora\_target} & "all" \\
        \texttt{template} & "qwen" \\
        \texttt{learning\_rate} & 1.0e-4 \\
        \texttt{num\_train\_epochs} & 3.0 \\
        \texttt{lr\_scheduler\_type} & "cosine" \\
        \texttt{warmup\_ratio} & 0.1 \\
        \hline
    \end{tabular}
    \caption{Parameter configuration for the fine-tuning of proxy model using LLaMAFactory.}
    \label{tab:llamafactory-params}
\end{table}

\paragraph{Evaluation Details.}
For each pair of (watermarking scheme, generation model), we construct the test set by sampling held-out instances from C4 and OpenGen that are disjoint from the training and validation splits. Specifically, we take the first 30 tokens of each instance as the prompt and generate both a watermarked and an unwatermarked continuation, forming paired test samples. In the main experiments, for each query text we use its first 50 tokens to generate a reference set of $N{=}16$ watermarked/unwatermarked pairs. For the Local Consistency Test module, we set the number of neighbors in the local support region to $k{=}7$. For the Adaptive Rank Test, we set $\lambda{=}0.6$ to balance overall fluency and local volatility. We report the main results averaged over five test sets, each containing 200 paired watermarked/unwatermarked samples.

\section{Settings for Robustness Evaluation}
\label{sec:appendix_robustness}

We evaluate robustness under six attacks, grouped into two categories: \emph{editing}-based transformations that directly modify the surface form, and \emph{paraphrasing}-based transformations that rewrite sentences more globally. All attacks are applied to the watermarked texts before running the detector.

\paragraph{Editing Attacks.} These editing attacks perturb the token sequence locally via deletion or substitution while largely preserving the overall semantics, with attack strengths controlled by the corresponding edit ratios.
\begin{itemize}[leftmargin=*]
    \item \textbf{Word-D} randomly deletes words with a deletion ratio of 0.3, namely each word is independently removed with probability 0.3. 
    \item \textbf{Word-S} performs synonym substitution based on WordNet~\cite{10.1145/219717.219748} with a substitution ratio of 0.5: it first identifies word positions that have available WordNet synonyms, then uniformly samples a subset (up to 50\% of all words) and replaces each selected word with a randomly chosen synonym.
    \item \textbf{Word-S-BERT} is a context-aware variant of synonym substitution. It samples word positions in the same manner as Word-S (ratio 0.5), masks each selected position, and uses a masked language model to predict a contextually plausible replacement token; positions where the mask token is truncated are skipped.
\end{itemize}

\paragraph{Paraphrasing Attacks.} These paraphrasing attacks rewrite the text more globally, inducing substantial lexical and/or syntactic variation while aiming to preserve the original semantics, thereby posing stronger distribution shifts than local edits.
\begin{itemize}[leftmargin=*]
    \item \textbf{Rephrase-Dipper-1} applies a restatement attack using the Dipper paraphraser~\cite{krishna2023paraphrasing}, emphasizing lexical variation while largely preserving sentence structure. We set $\texttt{lex\_diversity}=60$ and $\texttt{order\_diversity}=0$. 
    \item \textbf{Rephrase-Dipper-2} uses Dipper paraphraser with stronger rewriting level by enabling both lexical and word-order variation, producing more diverse restatements. We set $\texttt{lex\_diversity}=60$ and $\texttt{order\_diversity}=60$.
    \item \textbf{Rephrase-GPT5.1} paraphrases the text with OpenAI's GPT5.1 API~\cite{openai_gpt51_model_docs_2025} (temperature 0.2) instructed to rewrite the input while preserving its original meaning. The prompt for paraphrasing is shown in Figure~\ref{fig:prompt_gpt}.
\end{itemize}

\begin{figure}[!t]
  \centering
  \begin{tcolorbox}[
      width=\columnwidth,
      colback=gray!5,
      colframe=black!70,
      colbacktitle=black!80,
      coltitle=white,
      fonttitle=\bfseries,
      title=Rephrase-GPT5.1 Attack Prompt,
      rounded corners,
      boxrule=0.6pt
  ]
  {\small\ttfamily
  \textbf{Prompt:} Please rewrite the input paragraph while preserving its original meaning. Output only the rewritten paragraph.\textbackslash n\textbackslash n Input: \{text\}
  }
  \end{tcolorbox}
  \caption{Prompt used for the Rephrase-GPT5.1 paraphrasing attack.}
  \label{fig:prompt_gpt}
\end{figure}

\section{Impact of Text Length}
\begin{figure}[t]
  \centering
  \includegraphics[width=\linewidth]{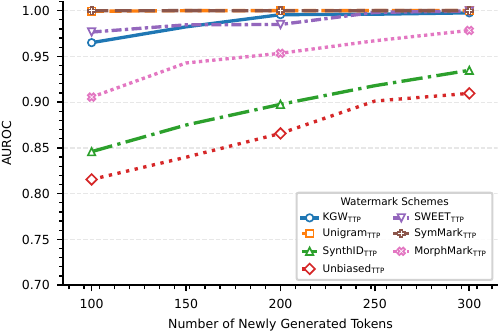}
  \caption{Impact of generated text length.}
  \label{fig:appendix_len}
\end{figure}
We further analyze how the generation length affects the detection performance of TTP-Detect. 
To eliminate length as a confounding factor, we always construct the reference set using completions whose length matches the query text, and then vary the number of newly generated tokens $L$ from 100 to 300. Figure~\ref{fig:appendix_len} reports the resulting AUROC scores.

Overall, longer texts consistently yield better detection across all watermark schemes, but the magnitude of improvement is highly scheme-dependent. 
For strong-signal schemes, performance saturates quickly: Unigram$_{\texttt{TTP}}$ and SymMark$_{\texttt{TTP}}$ remain near-perfect ($\approx 1.0$) even at $L=100$, while KGW$_{\texttt{TTP}}$ and SWEET$_{\texttt{TTP}}$ already exceed 0.95 at $L=100$ and essentially reach the ceiling by $L=250$. 
In contrast, schemes designed to be less distortive or more unbiased exhibit a much stronger dependence on text length.

This trend is consistent with the accumulation effect of token-level watermark signals: as $L$ increases, the expected watermark evidence grows while the variance of relative measurements (e.g., representation- or likelihood-based discrepancies against the reference set) decreases, making the watermarked and unwatermarked distributions more separable. 
Practically, these results suggest that moderate lengths (around 200 tokens) are sufficient for near-ceiling performance on strong schemes, whereas weaker-signal schemes may require longer generations to achieve comparable reliability—at the cost of increased generation and representation-extraction overhead.

\section{Hyperparameter Analysis}
\label{sec:appendix_hyperparameter}
\paragraph{Impact of $k$ in Local Consistency Test.}
In the Local Consistency Test, the hyperparameter $k$ specifies the number of nearest neighbors used to define the local support region around the query text's representation. Figure~\ref{fig:hyperparameter_k} reports the detection performance under $N{=}16$ while varying $k$. Across all seven watermarking schemes, the AUROC score improves as $k$ increases when $k<7$. When $k\ge 7$, the curves largely saturate and remain stable for most schemes. We therefore set $k{=}7$ as the default choice for $N{=}16$. For other reference set sizes, we follow a consistent heuristic and set $k$ to the largest odd number not exceeding $N/2$; for example, $k{=}15$ for $N{=}32$ and $k{=}31$ for $N{=}64$.

\paragraph{Impact of $\lambda$ in Adaptive Rank Test.}
\begin{figure}[t]
  \centering
  \includegraphics[width=\linewidth]{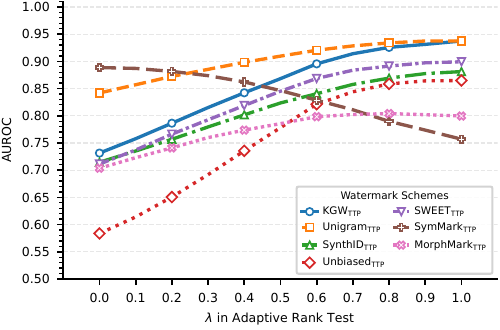}
  \caption{Hyperparameter analysis of $\lambda$ in Adaptive Rank Test module.}
  \label{fig:appendix_hyperparameter_lambda}
\end{figure}
In the Adaptive Rank Test, we use a mixing coefficient $\lambda\in[0,1]$ to balance two complementary signals: the overall likelihood shift and the local volatility. Figure~\ref{fig:appendix_hyperparameter_lambda} reports the resulting AUROC when sweeping $\lambda$ from 0 to 1, where larger $\lambda$ places more emphasis on the overall shift term.

Overall, most watermarking schemes benefit from increasing $\lambda$. In particular, KGW$_{\texttt{TTP}}$, SWEET$_{\texttt{TTP}}$, and SynthID$_{\texttt{TTP}}$ exhibit a largely monotonic improvement as $\lambda$ increases, suggesting that their detectable signal is dominated by a consistent global bias rather than highly irregular local variations. The effect is most pronounced for Unbiased$_{\texttt{TTP}}$, whose AUROC rises sharply from a weak regime at $\lambda=0$ to a competitive level at $\lambda\ge 0.6$, indicating that emphasizing the global shift is critical for revealing its more subtle watermark traces. By contrast, SymMark$_{\texttt{TTP}}$ shows the opposite trend: performance degrades steadily as $\lambda$ increases, implying that its signal is better captured by the volatility component and can be attenuated when the detector over-weights the global shift. MorphMark$_{\texttt{TTP}}$ shows moderate gains up to mid-range $\lambda$ and then saturates (or slightly drops), suggesting a weaker dependence on this trade-off.

Based on these observations, we set $\lambda=0.6$ globally as a robust compromise: it achieves near-saturated performance for the majority of schemes while avoiding overly aggressive emphasis on the global-shift term that would substantially harm SymMark$_{\texttt{TTP}}$. This choice provides a stable default without requiring per-scheme hyperparameter tuning.

\section{Full Result of Module Ablation}
\label{sec:module-ablation}
\begin{figure*}[!t]
  \centering
  \begin{subfigure}[t]{0.48\textwidth}
    \centering
    \includegraphics[width=\linewidth]{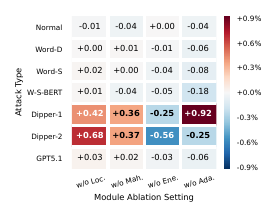}
    \caption{KGW$_\texttt{TTP}$}
  \end{subfigure}\hfill
  \begin{subfigure}[t]{0.48\textwidth}
    \centering
    \includegraphics[width=\linewidth]{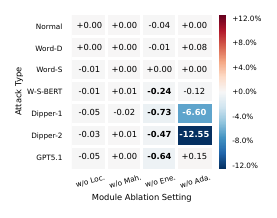}
    \caption{Unigram$_\texttt{TTP}$}
  \end{subfigure}
  \vspace{4pt}
  \begin{subfigure}[t]{0.48\textwidth}
    \centering
    \includegraphics[width=\linewidth]{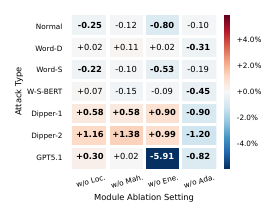}
    \caption{MorphMark$_\texttt{TTP}$}
  \end{subfigure}\hfill
  \begin{subfigure}[t]{0.48\textwidth}
    \centering
    \includegraphics[width=\linewidth]{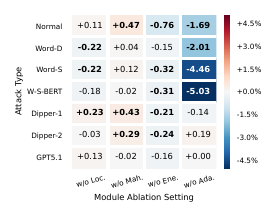}
    \caption{Unbiased$_\texttt{TTP}$}
  \end{subfigure}
  \vspace{4pt}
  \begin{subfigure}[t]{0.48\textwidth}
    \centering
    \includegraphics[width=\linewidth]{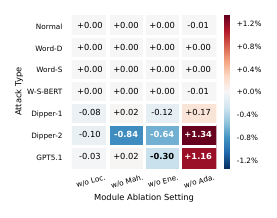}
    \caption{SymMark$_\texttt{TTP}$}
  \end{subfigure}\hfill
  \begin{subfigure}[t]{0.48\textwidth}
    \centering
    \includegraphics[width=\linewidth]{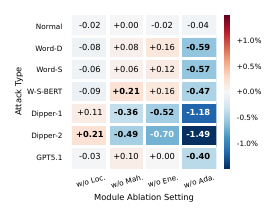}
    \caption{SWEET$_\texttt{TTP}$}
  \end{subfigure}
  \caption{Module ablation results for various watermarking schemes.}
  \label{fig:appendix_ablation_module}
\end{figure*}
We report the module-level ablation results for the remaining six watermarking schemes in Figure~\ref{fig:appendix_ablation_module}. For each scheme, we remove one of the four relative measurement modules at a time and re-evaluate the resulting detector under all attack settings. Each heatmap cell shows the change in AUROC relative to the full configuration, where darker colors indicate a larger impact. 

The contribution of each module varies across watermarking schemes, reflecting the joint effect of the watermarking scheme and the specific attack type, as schemes inject watermark signals with different structures and strengths that are disrupted differently by each attack. For instance, on Unigram$_\texttt{TTP}$, removing the Adaptive Rank Test causes an average AUROC decrease of 9.57\% under the two Dipper paraphrase attacks, whereas on Unbiased$_\texttt{TTP}$ the same ablation primarily affects editing-based attacks such as Word-D, Word-S, and Word-S-BERT. 

While we occasionally observe slight performance gains after removing a module for a specific attack, the overall ensemble remains effective: our weighting strategy is designed to improve average robustness across heterogeneous scenarios, rather than to over-optimize for any single setting.

\section{Practical Deployment Considerations}
\label{sec:appendix_deploy}

\subsection{Cooperation Among Parties and Minimal Trust Assumptions}
TTP-Detect is designed for a multi-party governance workflow, where watermark injection and verification are separated by design.
A feasible deployment requires coordination among (i) a regulator or standard-setting body $U$, (ii) a compliance-certified auditor $D$ (the trusted third party, TTP), and (iii) the model service provider $S$.
Concretely, the provider exposes a standard text-generation interface with a binary watermark-control flag (watermarked vs.\ unwatermarked), while keeping the watermark algorithm and secret key confidential.
The auditor only relies on observable outputs to (a) construct paired reference sets, (b) calibrate detection statistics, and (c) issue audit reports, without requiring access to the provider’s internal watermarking logic.
This separation of responsibilities enables neutral, third-party verification while minimizing trust assumptions: the provider controls injection, whereas verification and oversight are performed independently.

\subsection{Improving Reference Feasibility via Prompt Reconstruction}
\begin{figure*}[!t]
  \centering
  \begin{tcolorbox}[
      width=\textwidth,
      colback=gray!5,
      colframe=black!70,
      colbacktitle=black!80,
      coltitle=white,
      fonttitle=\bfseries,
      title=Prompt Reconstruction Instruction,
      rounded corners,
      boxrule=0.6pt
  ]
  {\small\ttfamily
  \textbf{Prompt:} You are given a response produced by a large language model, but the original user prompt is unknown. Infer a plausible prompt that could have elicited the response. Output \emph{only} the reconstructed prompt, with no explanation.\textbackslash n\textbackslash n Response: \{text\}
  }
  \end{tcolorbox}
  \caption{Reverse-prompting instruction used to reconstruct an unknown user prompt from the query text.}
  \label{fig:prompt_gpt_reconstruct}
\end{figure*}
In our experiments, reference texts are generated by prompting the provider $S$ with the prefix of the query text, producing paired watermarked/unwatermarked completions under the same prompt.
Although TTP-Detect can remain effective even without reference sets (see Sec.~\ref{sec:further-analysis}), tighter semantic alignment between the query and its references is generally beneficial for relative measurements, as it reduces semantic-induced variance that may confound watermark-induced discrepancies.

In real-world settings, however, prefix-only prompting may still yield semantic drift for certain queries (e.g., long-form or instruction-following outputs), because the original user instruction is unavailable to the auditor $D$.
As an optional enhancement, $D$ can reconstruct a plausible instruction that could have produced the query text, using an auxiliary LLM.
Given a received query text $t_{\text{q}}$, the auditor first infers an instruction $\hat{p}$ (``reverse-prompting''), and then queries the provider using a composite prompt formed by ``$\hat{p}$ + prefix($t_{\text{q}}$)'' to generate the paired reference set.
This lightweight procedure aims to make reference completions closer to $t_{\text{q}}$ in semantics and style, allowing relative measurement modules to focus on watermark differences rather than prompt mismatch.

In a pilot study on 50 paired KGW watermarked/unwatermarked samples constructed from ELI5~\cite{fan-etal-2019-eli5} using Llama-3.1-8B, we use the GPT-5.1 API as the auxiliary LLM for prompt reconstruction, which improves AUROC to 0.975 (a +7.7\% absolute gain over prefix-only prompting).
We emphasize that this result is preliminary: prompt reconstruction can be ambiguous and introduces additional computation, but it provides a practical path to improve reference feasibility when the original prompt is unavailable. The instruction for prompt reconstruction is shown in Figure~\ref{fig:prompt_gpt_reconstruct}.

\end{document}